\documentclass[structabstract]{aa}  
\usepackage{graphicx}
\def\arcsec{$^{\prime\prime}$}
\def\arcmin{$^{\prime}$}
\def\degrees{$^{\circ}$}
\begin{document}
   \title{The large-scale diffuse radio emission in A781}

   \subtitle{}

   \author{F. Govoni \inst{1},
           M. Murgia \inst{1},
           G. Giovannini \inst{2,3},
           V. Vacca  \inst{1,4}, \and 
           A. Bonafede \inst{5,3}
          }

  \institute{INAF - Osservatorio Astronomico di Cagliari,
             Strada 54, Loc. Poggio dei Pini, 09012 Capoterra (Ca), Italy
         \and  
             Dipartimento di Astronomia, Universit\`a degli Studi di Bologna, 
             Via Ranzani 1, 40127 Bologna, Italy
         \and             
            INAF - Istituto di Radioastronomia, Via P.Gobetti 101, 
            40129 Bologna, Italy 
         \and
              Dipartimento di Fisica, Universit\`a degli Studi di Cagliari, Cittadella Universitaria, I-09042 Monserrato (CA), Italy
         \and
              Jacobs University of Bremen, Campus Ring 1, 28759 Bremen, Germany
             }

   \date{Received September 15, 1996; accepted March 16, 1997}

  \abstract
  {A781 belongs to a complex system characterized by
    extended X-ray emissions that may form part 
   of line of clusters of galaxies along a filament.
   } 
   {The aim of this work is to investigate the possible presence of 
  extended, diffuse synchrotron radio emission connected to the intra-cluster medium of A781.
   }
   {We studied the radio continuum emission and the spectral index properties 
   in proximity of the A781 by analyzing archival Very Large Array observations
   at 1400 and 325 MHz.
   }
   {The main cluster of the system is permeated by diffuse 
   low-surface brightness radio emission which is classified, 
   being located close to the center, as a radio halo.
   The diffuse emission presents the typical 
   extension and radio power of the other halos known in the literature.
   Interestingly, the radio halo appears to be linked to 
   a peripheral patch previously found in the literature.
   The spectrum of this peripheral emission shows
   a radial steepening which may confirm that this source is indeed 
   a cluster relic. }
   {}

   \keywords{Galaxies:cluster:general  -- Galaxies:cluster:individual:A781  
    -- Magnetic fields -- (Cosmology:) large-scale structure of Universe}

   \titlerunning{The large-scale diffuse radio emission in A781}
   \authorrunning{F. Govoni et al.}
   \maketitle
%

\section{Introduction}

An ever increasing number of galaxy clusters exhibits at its
center a Mpc-scale synchrotron radio halo.
These elusive sources are characterized by a low-surface brightness 
($\sim 1 \mu$Jy/arcsec$^2$ at 1400 MHz) and steep-spectrum
\footnote{$S(\nu)\propto \nu^{- \alpha}$, with $\alpha$=spectral index}
($\alpha>1$). They represent one of the best evidence for 
the presence of relativistic electrons and magnetic fields 
in the intra-cluster medium.
 
In the last decade several projects aimed to detect new
radio halos have been performed (e.g. Giovannini et al. 1999,
Giovannini \& Feretti 2000, Kempner \& Sarazin 2001, 
Govoni et al. 2001a, Bacchi et al. 2003,
Venturi et al. 2007, Venturi et al. 2008, van Weeren et al. 2009, 
Rudnick \& Lemmerman 2009, Giovannini et al. 2009).
The investigation on the physical
properties of radio halos and their hosting environments have improved our
knowledge and led to the formulation of possible scenarios
of their formation, which are still matter of debate 
(e.g. Brunetti et al. 2009, En{\ss}lin et al. 2010). 
The radio halo morphology is often very similar to the X-ray emitting
thermal intra-cluster medium (e.g. Govoni et al. 2001b,
Feretti et al. 2001, Giacintucci et al. 2005).
Furthermore, radio halos are preferentially found in clusters 
showing evidence of merger activity (e.g. Buote 2001, Schuecker et al. 2001,
Govoni et al. 2004, Cassano et al. 2010),
 suggesting a connection between the origin of radio halos 
and gravitational processes of cluster formation, 
although a one-to-one association between 
merging clusters and radio halos is not supported by present observations. 
It is thus fundamental in this context to better investigate the radio emission from merging and X-ray luminous galaxy clusters.

As part of an ongoing program aimed to investigate in complex X-ray cluster systems 
the presence of halo emission,
we recently found (Murgia et al. 2010) the first example of a 
double radio halo in the close pair of 
galaxy clusters A399 and A401.
Motivated by the above discovery, we analyzed archival Very Large Array (VLA)
radio observations of A781.
This is an exceptional system, since combined X-ray and 
weak-lensing analysis (Wittman et al. 2006, Sehgal et al. 2008) 
indicate that A781 is really a complex of several clusters.
 
In this work we show how this system appears at the radio wavelengths.
The paper is organized as follows: 
In Sect. 2 we describe the properties of the cluster.
In Sect. 3 we present the radio observations, and the data reduction. 
In Sect. 4 we show the results of the total intensity and spectral index
images and finally, in Sect. 5 we draw the conclusions.

Throughout this paper we assume a $\Lambda$CDM cosmology with
$H_0= 71~\mathrm{ km\, s^{-1}\, Mpc^{-1}}$,
$\Omega_m=0.27$, and $\Omega_{\Lambda}=0.73$. At the 
distance of A781 (z=0.3004, Geller et al. 2010), 1\arcsec~corresponds 
to 4.4 kpc.


\section{The A781 system}

\begin{figure}[t]
\centering
\includegraphics[width=9 cm]{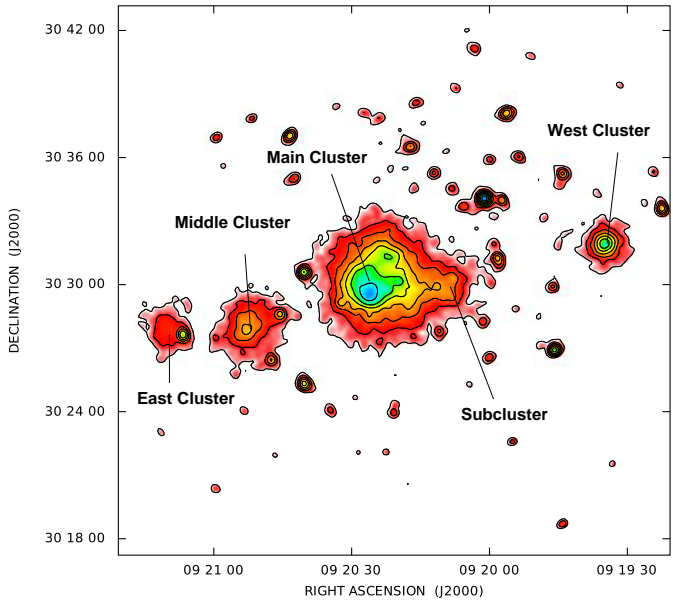}
\caption{
XMM-Newton image, in the 0.2-12 keV band, of the A781 cluster complex.
The X-ray image has been convolved with a Gaussian 
of $\sigma$=8\arcsec.
The first contour level is drawn at 3 cts/Skypixel (1~Skypixel=4$''$)
and the rest are spaced by a factor of $\sqrt 2$.
The clusters are labelled as in Sehgal et al. (2008).
}
\label{xmm}
\end{figure}

A781 belongs to a complex system characterized by
extended X-ray sources that may form part 
of line of clusters along a filament (e.g. Jeltema et al. 2005).
The X-ray surface brightness distribution
of the system is shown in Fig.~\ref{xmm}, where we
present an image in the 0.2$-$12 keV band obtained with a recent 
XMM-Newton (MOS1+MOS2) exposure time of 80 ksec (Obs. id: 0401170101).
As previously pointed out by Sehgal et al. (2008), the X-ray 
image reveals that A781 consists of a large 
main (``Main'') cluster connected to a sub-cluster.
In addition, two smaller clusters to its east (``Middle'' and ``East''), 
and one to its west (``West'') are in seen at a relatively small 
separation on the sky.
The X-ray and weak-lensing analysis by Sehgal et al. (2008) suggests 
that the ``Main'' cluster may be undergoing a merger 
with the ``Subcluster'' to southwest, while the other 
clusters are rather relaxed.
We note that the redshift of the ``Middle'' cluster (z=0.2915) is quite similar to that of the
``Main'' cluster (z=0.3004). On the other hand, the ``East'' and the ``West'' clusters 
have different redshifts (z=0.4265 and z=0.4273 respectively; 
Geller et al. 2010), 
therefore they are not related to the main cluster of the system.

The A781 system was observed with the 
Giant Metrewave Radio Telescope at 610 MHz by Venturi et al. (2008).
An interesting peripheral patch of diffuse emission,
with no obvious optical counterpart, peaking at R.A.(J2000)=09h20m32.2s 
Decl.(J2000)=30\degrees 27\arcmin 34.2\arcsec~ has been detected, 
and the authors suggested that it might be a relic source, but
no hint of radio halo emission was found.

The presence of diffuse cluster-scale radio emission close to the
``Main'' cluster X-ray centroid, indicative of a radio halo, 
was suggested by Rudnick \& Lemmerman (2009), by
reprocessing radio images from the Westerbork Northern Sky Survey 
(WENSS, Rengelink et al. 1997) at 327 MHz.

\section{VLA observations and data reduction}

\begin{table*}
\caption{Details of the VLA observations.}
\begin{center}
\begin{tabular} {lccccclll} 
\hline
Source   &  R.A.      &  Decl.       &  Frequency      & Bandwidth   &  Conf. &  Time    & Date & Program\\
         &  (J2000)   &  (J2000)     &  (MHz)          & (MHz)       &        &  (minutes) &      &         \\
\hline

A781     & 09 20 23.7 & $+$30 31 09.0 &  1365/1435     &  7$\times$3.125 &  A  & 30            & 1994 April 19 & AB699  \\
         & 09 20 22.6 & $+$30 31 20.6 &  1452/1502     &  25             &  C  &  8            & 1984 May 4    & AO048 \\
         & 09 20 23.7 & $+$30 31 09.0 &  1365/1435     &  7$\times$3.125 &  D  & 15            & 1995 March 15 & AM469\\
         & 09 14 36.0 & $+$29 44 52.0 &  327.5/321.6   &  64$\times$0.048  &  B  & 49          & 2005 May 21                 & AD509        \\
         & 09 14 36.0 & $+$29 44 52.0 &  327.5/321.6   &  64$\times$0.048  &  C  & 183,~17     & 2005 Sept. 29, Oct. 3 & AD509   \\
         & 09 14 36.0 & $+$29 44 52.0 &  327.5/321.6   &  64$\times$0.048  &  D  & 88,~96,~20  & 2005 Dec. 4,~9,~12        & AD509   \\
\hline
\multicolumn{8}{l}{\scriptsize Col. 1: Source; Col. 2,3: Pointing position (R.A., Decl.); Col. 4: Observing frequency; 
                   Col. 5: Observing Bandwidth;}\\
\multicolumn{8}{l}{\scriptsize Col. 6: VLA configuration; Col. 7: Time on source; 
Col. 8: Dates of observation; Col. 9: VLA program.}\\
\end{tabular}
\label{tab1}
\end{center}
\end{table*}

Because of the peculiar X-ray morphology of this complex system 
and previous radio results, we investigated the radio 
continuum emission in proximity of the 
A781 system at 1400 and 325 MHz.
The details of the radio observations are given in Table \ref{tab1}.
Calibration and imaging were performed with the
NRAO Astronomical Image Processing System (AIPS) package.

\subsection{Observations at 1400 MHz}

We analyzed archival VLA observations
at 1400 MHz in A, C, and D configuration.
The data were calibrated in phase and amplitude.
Data editing has been made in order to excise 
radio frequency interferences (RFI).
In the A and D configuration the observations were done in  
spectral line mode therefore, the data were calibrated in bandpass  
and the data editing was done channel by channel.
Surface brightness images were produced
following the standard procedures: Fourier-Transform,
Clean, and Restore implemented in the 
AIPS task IMAGR. We used the Multi-scale CLEAN 
(see e.g. Greisen et al. 2009), 
an extension of the classical Clean algorithm,
implemented in the task IMAGR. 
We averaged the 2 IFs 
(and the 7 channels in the D and A configuration) 
together in the gridding process under IMAGR. 
Self-calibration (phase) was performed to increase the
dynamic range and sensitivity of the radio images.

A polarization sensitive image has been produced by using
the C configuration data-set (the spectral line mode of the 
A and D configuration is not suitable for this purpose).

\subsection{Observations at 325 MHz}

There are no pointed low frequency VLA observations of A781. 
However, the cluster falls within the field of view 
(at a distance of $\simeq$1.5\degr~from the pointing) of
the project AD509, which consists of archival observations
at 325 MHz in B, C, and D configuration.
The data were collected in spectral line mode with a total bandwidth 
of 3.1 MHz, subdivided into 64 spectral channels.  
The flux density scale and the bandpass were calibrated by using 
3C48 (3C147 in B configuration).
The source 0909+428 was used as phase calibrator.
Task FLGIT was applied to automatically remove data with 
strong RFI  
(about 20-30\% of the data were typically flagged). 
The data was averaged to 10 channels (with a width of $\simeq$0.3 MHz),
keeping the effects of bandwidth smearing under control.
Low-level residual RFI were carefully removed from the 10 channels 
data-set by visually inspection and finally the data were extracted 
with task SPLIT and imaged.  
We processed the data-sets separately applying several cycles of imaging and self-calibration, to remove residual phase 
variations. All the data-sets were then combined together with task DBCON and we performed a final self-calibration run.
The area was imaged in facets (Cornwell \& Perley 1992) of  
slightly overlapping fields in order to account for the non-coplanarity 
of the incoming wavefront within the large primary beam of $\sim 3\degr$.
The deconvolution was performed with a Multi-scale CLEAN. 

\section{Results}

We analyze the results of the VLA observations with a particular emphasis
on the analysis of the radio properties 
of the large-scale diffuse cluster emission. 

\subsection{Cluster radio emission at 1400 MHz}

\begin{figure*}[t]
\centering
\includegraphics[width=18 cm]{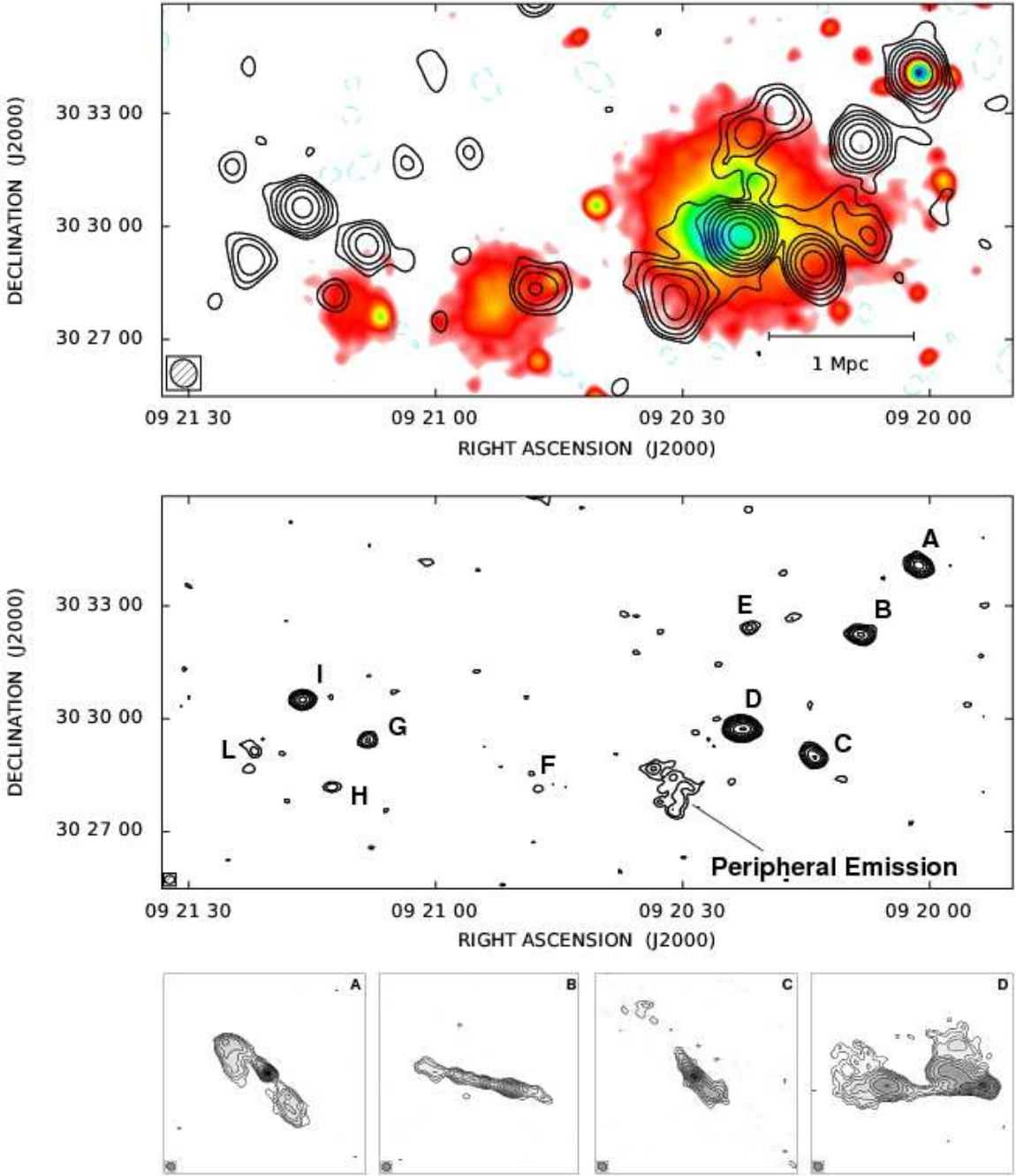}
\caption{
Top: Total intensity radio contours of the A781 system at 1400 MHz
with the VLA in D configuration.
The image has a FWHM of 41.6\arcsec$\times$44.8\arcsec (PA=16.6\degr).
The contour levels are drawn at $-$0.3 (thin-dashed lines), 0.3 mJy~beam$^{-1}$,
and the rest are spaced by a factor of 2.
The sensitivity (1-$\sigma$) is 0.1 mJy~beam$^{-1}$.
Total intensity radio contours are overlaid on the XMM X-ray 
image in the 0.2-12 keV band. 
The X-ray image has been convolved with a Gaussian of $\sigma$=8\arcsec.
Middle: Total intensity radio contours at 1400 MHz
with the VLA in C configuration.
The image has a FWHM of 13.2\arcsec$\times$16.5\arcsec (PA=$-$84.9\degr).
The first contour level is drawn at 0.27 mJy~beam$^{-1}$
and the rest are spaced by a factor of 2.
The sensitivity (1-$\sigma$) is 0.09 mJy~beam$^{-1}$.
Bottom: Total intensity radio contours of the discrete
sources A, B, C, and D at 1400 MHz with the VLA in A configuration.
The images have a FWHM of 1.2\arcsec$\times$1.4\arcsec (PA=67.3\degr).
The sensitivity (1-$\sigma$) is 0.05 mJy~beam$^{-1}$.
The first contour level is drawn at 0.15 mJy~beam$^{-1}$
and the rest are spaced by a factor of $\sqrt2$.
The field of view of the images is 30\arcsec.
}
\label{radio}
\end{figure*}

The radio iso-contours at 1400 MHz of the ''Main'', ``Middle'' and ``East''
cluster of the system are shown in the top panel of Fig.~\ref{radio}.
This image has been obtained with the VLA in D configuration
and has a FWHM beam of 41.6\arcsec$\times$44.8\arcsec.
In order to compare radio and X-ray cluster emission, 
the radio iso-contours are overlaid on the XMM-Newton image 
presented in Fig.~\ref{xmm}. 
We find that the central region of the ``Main'' cluster is permeated by   
diffuse low-surface brightness emission which we classified as a radio halo.
The radio halo appears to be linked to 
the brighter peripheral patch 
previously found by Venturi et al. (2008).
In addition, the diffuse radio emission of the ``Main'' cluster 
is elongated toward the ``Subcluster''. 
As measured from the 3-$\sigma$ radio isophote, the overall diffuse 
emission has an angular extension 
of about 7\arcmin~($\simeq 1.8$ Mpc at the cluster distance).

\begin{figure*}[t]
\centering
\includegraphics[width=18 cm]{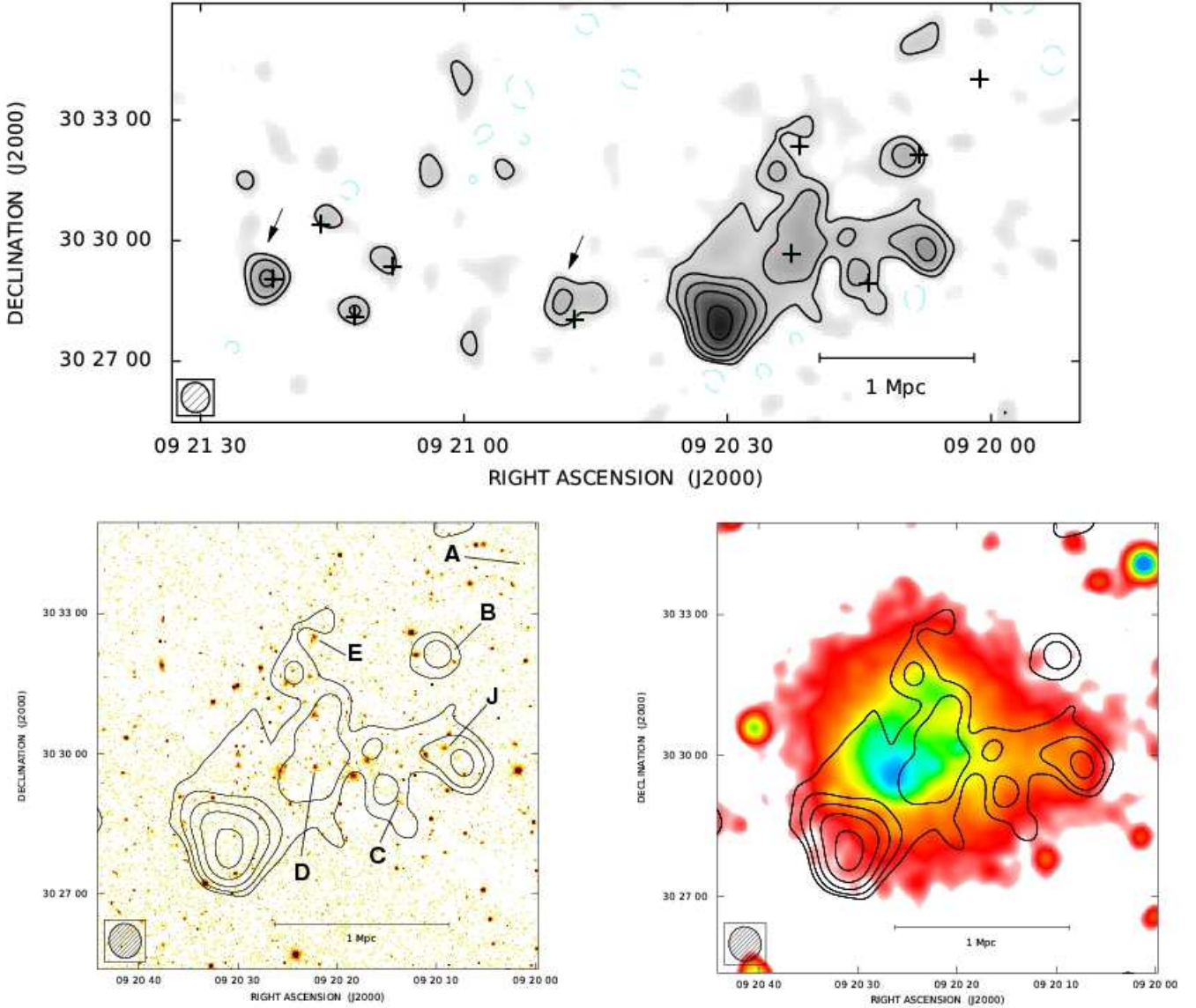}
\caption{
Top: Total intensity radio contours at 1400 MHz with the VLA in D configuration
after subtraction of discrete sources.
The image has a FWHM of 42\arcsec$\times$45\arcsec (PA=10.3\degr).
The contour levels are drawn at $-$0.3 (thin-dashed lines), 0.3 mJy~beam$^{-1}$,
and the rest are spaced by a factor of 2.
The sensitivity (1-$\sigma$) is 0.1 mJy~beam$^{-1}$.
Crosses indicate the positions of the subtracted discrete sources. 
Arrows indicate hints of possible diffuse emission
left after the subtraction process.
Bottom (left): Zoom of the ``Main'' cluster in which the total 
intensity radio contours are overlaid on the red plate of the
Sloan Digital Sky Survey.
Bottom (right) : Zoom of the ``Main'' cluster in which the total 
intensity radio contours are overlaid on the XMM X-ray image.
}
\label{radio_sub}
\end{figure*}

To separate the diffuse radio emission from discrete sources 
we produced images at higher resolution.
In the middle panel of Fig.~\ref{radio} we present the radio 
iso-contours of A781, 
taken with the VLA in C configuration.
This image has a FWHM beam of 13.2\arcsec$\times$16.5\arcsec.
Although the relatively high resolution of this image is not
particularly suitable to detect diffuse cluster emission, 
the peripheral patch of diffuse emission is clearly 
visible in the C array data-set too.
The discrete sources are labelled in the figure and their
position and flux densities are given in Table \ref{tab2}.
All of them show an optical counterpart in the Sloan Digital Sky
Survey (see bottom left panel of Fig.~\ref{radio_sub}). 
The only exception is source B, which may be a background radio galaxy.
Source D, which is located about 50\arcsec~west to the X-ray peak 
of the ``Main'' cluster, is the brightest radio source
in the field.
Its fractional polarization is $\simeq$ 3\%.
The other discrete sources embedded in the 
halo have a fractional polarization below the 3-$\sigma$ noise level. 
The peripheral diffuse emission appears unpolarized too and
given that its surface brightness is $\simeq$0.5 mJy~beam$^{-1}$, and the sensitivity 
(1-$\sigma$) of the linear polarization image
is $\simeq$0.08 mJy~beam$^{-1}$, we can 
just set a loose 3-$\sigma$ upper 
limit to the fractional polarization of $<$48\%, consistent with 
the values found for other relic sources.

The image obtained with the VLA in A configuration has a  
FWHM beam of  1.2\arcsec$\times$1.4\arcsec~and, in agreement
with the C array configuration, confirms that just a few discrete
sources are present in the field. 
Among the discrete sources, those labelled with A, B, C and D 
(shown in the bottom 
panel of Fig.~\ref{radio}) are extended. 
While A, C, and D show the typical morphology of ``head-tail'' 
radio sources in cluster, source B shows a straight ``naked-jet''
morphology.

The total flux density in the region of 
the ``Main'' and ``Subcluster'' is $\simeq$ 119$\pm$4 mJy.
By subtracting the flux density 
of the embedded discrete sources C, D, and E as derived in the
C configuration data-set (see Table \ref{tab2}),
a flux density of $\simeq$ 36$\pm$5 mJy appears to be associated 
with the low brightness diffuse emission.
This flux density value corresponds to a radio power 
$P_{\rm 1400 MHz} = 1.0\times10^{25}$ W~Hz$^{-1}$. 
A781 is part of the ROSAT Brightest Cluster sample by Ebeling et al. (1998). 
Its X-ray luminosity in the 0.1-2.4 keV band
is $1.1\times 10^{45}$ erg/sec. 
The radio power $P_{\rm 1400 MHz}$, the radio largest linear size ($LLS$),  
and the X-ray luminosity ($L_X$), are in
agreement with the $P_{\rm 1400 MHz}-LLS$ and $P_{\rm 1400 MHz}-L_X$ relations 
known for the other halos in clusters.
The peripheral patch alone has a flux density $\simeq$15.5$\pm$0.5 mJy.
Thus, if we consider the radio halo separated by 
the peripheral patch it results in a flux density 
of $\simeq$20.5$\pm$5 mJy. This value correspond to 
a radio power $P_{\rm 1400 MHz} = 5.9\times10^{24}$ W~Hz$^{-1}$ ,
still in agreement with the $P_{\rm 1400 MHz}-LLS$ and $P_{\rm 1400 MHz}-L_X$ 
relations known in the 
literature (see e.g. Giovannini et al. 2009).

The flux densities were calculated, after the primary beam correction, 
by integrating the total intensity surface brightness, down to the 3$\sigma$ 
level.
However, we note that the residual flux density associated with
the diffuse cluster emission must be interpreted with caution
because of a possible variation in the discrete sources flux density, 
the slightly different frequency of the two data-sets, 
and any absolute calibration error between 
the two data-sets could cause in an under or 
over subtraction of flux. 
In addition, due to the short exposure time of the archival observations,
some diffuse emission could be missed. 
Therefore, a deep follow-up investigation would be necessary 
to ensure to recover all the radio flux and to unambiguously 
separate the emission of the diffuse emission from that 
of the unrelated discrete sources.

To ensure that the large-scale diffuse emission
is not due to the blending of discrete sources,  we present the total intensity 
radio contours at 1400 MHz with the VLA in D configuration
in the top panel of Fig.~\ref{radio_sub},
after subtraction of discrete sources.
To obtain it, we produced an image of the discrete 
sources by using only the longest baselines of the 
D configuration data-set, and uniform weighting.
The clean components of this image were then subtracted 
in the uv-plane by using the AIPS task UVSUB.
The image with the discrete sources subtracted confirms the presence
of a low-surface brightness radio halo at the cluster center 
connected to a brighter patch of peripheral emission to the south-east.

In the bottom panel of Fig.~\ref{radio_sub} 
we show a zoom of the ``Main'' cluster  in which the total 
intensity radio contours, after subtraction of discrete sources, 
are overlaid on the red image of the Sloan Digital Sky Survey (left)
and on the XMM X-ray image (right).
In this figure it is further clear that the elongation 
of the radio emission versus west
is located in coincidence with the ``Subcluster''.
We note that this feature is visible at 1400 MHz 
in the D configuration data-set only, while in the higher 
resolution images presented here, and in the FIRST image 
(Becker, White \& Helfand 1995), any discrete source seems
to be present.
On the other hand, in this location a source, classified
as a discrete source, is detected in GMRT images 
at 610 MHz (Venturi et al. 2008) and at 327 MHz (Giacintucci et al. 2010).

We note that, hints of possible diffuse emission 
(indicated by arrows in the top panel of Fig.~\ref{radio_sub}),
left after the subtraction process, are present on east of the ``Main''
cluster.
These diffuse radio emissions might trace the process 
of a large-scale structure formation, 
where cosmic shocks originated by complex 
merger events are able to amplify magnetic fields and accelerate 
synchrotron electrons along a cluster filament,
although a deeper observation 
is required to confirm the presence 
of these faint emissions.

\begin{table}
\caption{Information on discrete radio sources.}
\begin{center}
\begin{tabular} {lccl} 
\hline
Label   &  R.A.      &  Decl.       &  S$_{\rm 1400 MHz}$      \\
               &  (J2000)   &  (J2000)     &  (mJy)       \\
\hline
A              & 09 20 01.3 & 30 34 05     &  21.1$\pm$ 0.6     \\
B              & 09 20 08.3 & 30 32 15     &  14.9$\pm$ 0.5      \\
C              & 09 20 14.0 & 30 29 00     &  14.7$\pm$ 0.5      \\
D              & 09 20 22.7 & 30 29 45     &  67.0$\pm$ 2.0      \\
E              & 09 20 21.9 & 30 32 25     &   1.5$\pm$ 0.05     \\
F              & 09 20 48.6 & 30 28 36     &   1.0$\pm$ 0.03       \\
G              & 09 21 08.3 & 30 29 26     &   3.9$\pm$ 0.1  \\
H              & 09 21 12.7 & 30 28 11     &   1.5$\pm$ 0.05 \\  
I              & 09 21 16.2 & 30 30 31     &  20.7$\pm$ 0.6 \\
L              & 09 21 22.2 & 30 29 11     &   2.5$\pm$ 0.08   \\ 
\hline
\multicolumn{4}{l}{\scriptsize Col. 1: Source label; Col. 2, 3: Source position}\\ 
\multicolumn{4}{l}{\scriptsize (R.A., Decl.); Col. 4: Source flux density at
1400 MHz.}\\
\end{tabular}
\label{tab2}
\end{center}
\end{table}

\subsection{Cluster radio emission at 325 MHz}

\begin{figure*}[t]
\centering
\includegraphics[width=18 cm]{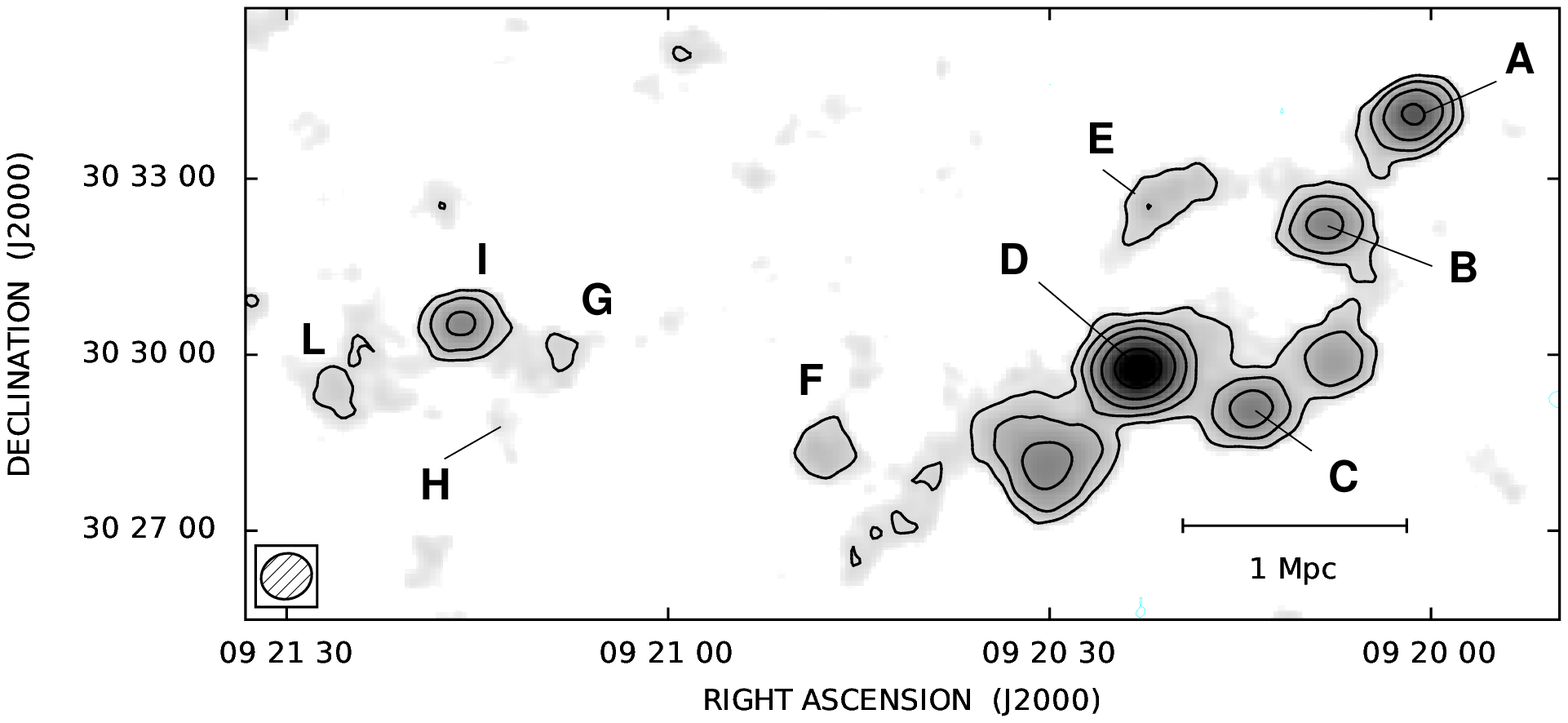}
\caption{
Total intensity radio contours of the A781 system at 325 MHz
with the VLA in B+C+D configuration.
The image has a FWHM of 46.1\arcsec$\times$52.0\arcsec (PA=$-$71.8\degr).
The contour levels are drawn at $-$2.9 (thin-dashed lines), 2.9 mJy~beam$^{-1}$,
and the rest are spaced by a factor of 2.
The sensitivity (1-$\sigma$) of this image (not corrected for the primary beam) is 0.95 mJy~beam$^{-1}$.
The image primary beam corrected has a sensitivity (1-$\sigma$) of 2.2 mJy~beam$^{-1}$.
}
\label{radio2}
\end{figure*}

In Fig.~\ref{radio2} we show the radio iso-contours of A781 at 325 MHz.
This image has been obtained by combining the VLA data in B, C, 
and D configuration. The resulting image has a
FWHM beam of 46.1\arcsec$\times$52.0\arcsec.
The VLA image has an angular resolution similar to the 
tapered GMRT image at 327 MHz presented by Giacintucci et al. (2010),
and displays the same structures.

Most of the features visible at 1400 MHz are also present  
at 325 MHz. In particular, the south-east peripheral patch and all the 
discrete sources, with the only exception of the source H, 
are clearly detected.
There is a hint of diffuse emission on the right of source D,
although most of the radio halo emission visible at 1400 MHz is missing 
at 325 MHz, likely because of the lower sensitivity 
of the low-frequency image. The only feature which appears to 
be comparatively brighter at 325 MHz
is the emission in coincidence with the ``Subcluster''. 
As we will see in the next Section, this feature is characterized by 
a very steep radio spectrum. 

By excluding the peripheral patch, the radio halo
has a flux density at 1400 MHz of S$_{\rm 1400 MHz}\simeq$20.5 mJy.
In the same area of about 1Mpc$^2$, the upper limit to the flux density at
325 MHz is S$_{325 MHz}<$137 mJy. 
This limit has been calculated by considering that 
the surface brightness of the halo, in the primary beam corrected image,
is everywhere lower than the 3$\sigma$ noise level (i.e. 6.6 mJy~beam$^{-1}$). 
Thus, we derive an upper limit to the global halo spectral index of 
$\alpha_{tot}<$1.3.

\subsection{Spectral index analysis}

In this section we present the spectral index image 
of A781 between 325 MHz and 1400 MHz. In order to be properly compared, 
the two images have been corrected for the primary beam 
attenuation of the VLA antennas, regridded to the same geometry, and
convolved to a common resolution of 53\arcsec$\times$53\arcsec.
We stress, however, that the original resolutions of the two images
were already very close not just due to the tapering of the uv-data
but because of the very similar intrinsic coverage of the relevant 
spatial frequencies.
We do not present the spectral index image after subtraction 
of discrete sources because of the low sensitivity of the 325 MHz 
image.

\begin{figure*}[t]
\centering
\includegraphics[width=18 cm]{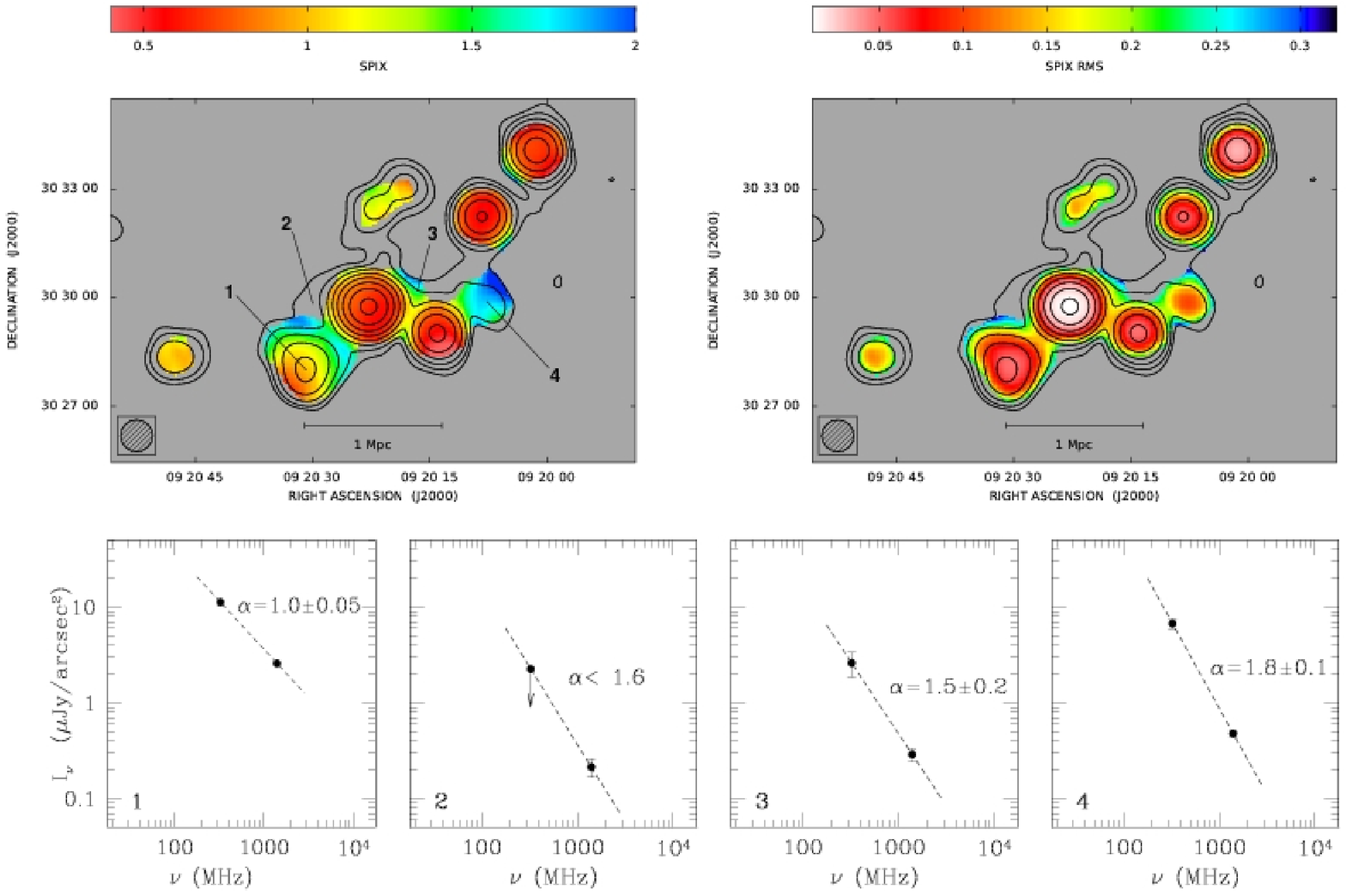}
\caption{
Top left: Spectral index image between 325 and 1400 MHz. The image has 
a FWHM of 53\arcsec $\times$ 53\arcsec. The 
iso-contours represent the 1400 MHz image convolved to a resolution 
of 53\arcsec $\times$ 53\arcsec and corrected for 
the primary beam attenuation.
The first contour level is drawn at 0.45 mJy~beam$^{-1}$
and the rest are spaced by a factor of 2. The 
sensitivity (1-$\sigma$) is 0.15 mJy~beam$^{-1}$.
The image at 325 MHz, convolved and primary beam corrected,
has a sensitivity (1-$\sigma$) of 2.5 mJy~beam$^{-1}$.
Top right: Image of the spectral index uncertainty. 
Bottom: Spectrum between 325 MHz and 1400 MHz in the four sample 
positions indicated in the top right panel.
}
\label{spix}
\end{figure*}

In the top panel of Fig.~\ref{spix}, we present 
the spectral index (left) and the spectral index 
uncertainty (right) images 
between 325 and 1400 MHz with the 1400 MHz radio 
iso-contours (primary beam corrected) overlaid.
They are calculated only on those pixels whose brightness
is above the $3\sigma$ level at both frequencies.
The spectral index values ranges between $\alpha \simeq$0.5 
and $\alpha \simeq$2, while the corresponding 
errors are in the range $\simeq$0.02$-$0.25.
The discrete sources have a typical spectral index value of  
$\alpha\simeq 0.6-0.7$, only the source F has a steep
spectrum with $\alpha$=1.0$\pm$0.2. 
In the bottom panel of Fig.~\ref{spix} we show the
spectrum between 325 MHz and 1400 MHz in four sample 
positions.

The spectral index of the diffuse emission is clearly determined mainly  
in coincidence with the peripheral patch and
the ``Subcluster'' (see positions 1 and 4 in Fig.~\ref{spix}).
The peripheral patch shows a clear spectral index trend
from  $\alpha \simeq$0.9$\pm$0.1, in the south-est tip, 
to $\alpha \simeq$1.6$\pm$0.2, close to the source D.
The radio elongation in coincidence with the 
``Subcluster'' has an extremely steep spectrum with 
$\alpha \simeq$1.8$\pm$0.1, which gives indication about the nature 
of this emission (see Sect. 5.2).

\section{Discussion and Conclusion}
Confirming the suggestion by Rudnick \& Lemmerman (2009),
we find that the central region of the ``Main'' cluster of A781 
is permeated by diffuse low-surface brightness 
radio emission, which is classified as a radio halo.
The radio halo is clearly detected at 1400 MHz while most
of its emission is missing at 325 MHz, likely because of the 
lower sensitivity of the low-frequency image.
At 1400 MHz, the diffuse emission presents the typical 
extension and radio power of the other halos known in literature.
In addition, two interesting features are found:\\ 
1) the radio halo is spatially connected to a 
peripheral diffuse patch, previously 
discovered by Venturi et al. (2008);\\
2) the cluster radio emission shows an 
elongation toward  the ``Subcluster''.

\subsection{The peripheral diffuse patch}
The peripheral patch may be interpreted as part of the halo or as
a radio relic (as suggested by Venturi et al. 2008
and Giacintucci et al. 2010). Radio relics are commonly defined as elongated 
diffuse cluster radio emission not associated with the 
cluster center (e.g Giovannini \& Feretti 2004). 
They are usually interpreted as linked to the presence of shock waves
that propagate in the intra-cluster medium during the 
process of cluster formation 
(e.g. En{\ss}lin et al. 1998, En{\ss}lin \& Gopal-Krishna 2001, 
Hoeft \& Br{\"u}ggen 2007).
Relics formation models
predict a steepening of the radio spectral index toward the cluster 
center, and a high degree of polarization 
(e.g. Clarke \& En{\ss}lin 2006, Bonafede et al. 2009, 
van Weeren et al. 2010).
In the case of the peripheral patch emission of A781,
the poor upper limit in the fractional
polarization does not permit to derive useful information.
However, its location and its radio
spectrum, which steepens toward the cluster center,
may favour a relic origin for this emission.

\subsection{The radio emission of the ``Subcluster''}
The diffuse radio emission of the ``Main'' cluster 
is elongated toward the ``Subcluster'', 
in a very similar way as the X-ray emission. 
The spectral index image reveals that the right end of this elongation
has a very steep spectral index of 
$\alpha \simeq$1.8$\pm$0.1. This emission seems to be really diffuse at
1400 MHz, since no point source is detected in this position neither
in the C array nor in the A array image.
Since this extreme
steep spectrum is not compatible with an active galaxy, 
a possibility is that this may be a dying radio 
galaxy (e.g. Murgia et al. 2011).
If this is the case, the optical progenitor could be 
the galaxy labelled J in the
bottom left panel of Fig.~\ref{radio_sub}. 
Venturi et al. (2008) found a faint radio counterpart
associated to this galaxy with the GMRT at 610 MHz.
Note, however, that the peak of the emission at 1400 MHz
is not precisely coincident with the galaxy J. 
Indeed, another possibility is that this emission is part
of the radio halo or even a steep spectrum halo 
connected with the ``Subcluster''. 

\subsection{Tracing the diffuse radio emission along a cluster filament}
In addition to the diffuse radio emission related to the ``Main'' 
and the ``Subcluster'',
on a larger scale, there are also some hints of diffuse emission 
elongated versus east.
The nature of these faint patches, which are
left after the subtraction of the discrete sources, is however rather uncertain
in view of the current data.
Nevertheless, if confirmed,
the presence of these patches of diffuse emission
may trace the process of a large-scale
structure formation, where cosmic shocks originated by complex 
merger events are able to amplify magnetic fields and accelerate 
synchrotron electrons along a large-scale filament of the cosmic web.

\begin{acknowledgements}
We thank the anonymous referee for the suggestions that 
improved the presentation of the paper, and Luigina Feretti for the support 
and the helpful discussions.
This research was partially supported by PRIN-INAF 2009.
The National Radio Astronomy Observatory (NRAO) is a facility 
of the National Science Foundation, operated under
cooperative agreement by Associated Universities, Inc. 
Funding for the SDSS and SDSS-II has been provided by the Alfred P. Sloan Foundation, the Participating Institutions, the National Science Foundation, the U.S. Department of Energy, the National Aeronautics and Space Administration, the Japanese Monbukagakusho, the Max Planck Society, and the Higher Education Funding Council for England. The SDSS Web Site is http://www.sdss.org/.
This research made use of Montage, funded by the National Aeronautics and Space Administration's Earth Science Technology Office, Computational Technnologies Project, under Cooperative Agreement Number NCC5-626 between NASA and the California Institute of Technology. The code is maintained by the NASA/IPAC Infrared Science Archive
\end{acknowledgements}

\end{document}